\documentstyle[twocolumn,aps,prl]{revtex}
\input epsf
\begin{document}
\def\beq{\begin{equation}}
\def\eeq{\end{equation}}
\def\beqn{\beqin{eqnarray}}
\def\eeqn{\end{eqnarray}}
\title{Thermodynamics of the Superconducting Phase Transition in  
Ba$_{0.6}$K$_{0.4}$BiO$_3$}
\author{P. Kumar} 
\address{Department of Physics, University of Florida, Gainesville,
FL  32611}
\author{Donavan Hall} 
\address {National High Magnetic Field Laboratory, 
Florida State University, Tallahassee, FL 32310}
\author{R.G. Goodrich} 
\address {Department of Physics and Astronomy, Louisiana State
University, Baton Rouge, LA  70803--4001}

\date{\today}
\maketitle

\widetext

\begin{abstract}
We suggest that the transition to superconductivity in a single 
crystal of Ba$_{0.6}$K$_{0.4}$BiO$_3$ with a $T_c$ = 32K, and having 
critical fields with anomalous temperature dependencies and vanishing 
discontinuities in specific heat and magnetic susceptibility, may well 
be an example of a fourth order (in Ehrenfest's sense) phase 
transition.  We have derived a free energy functional for a fourth 
order transition and calculated (for the temperature range ${T_c/2 < T 
\simeq T_c}$) the temperature dependence of the critical fields.  We 
find $H_{c1}(T)\propto (1-T/T_c)^3$, $H_0(T)\propto (1-T/T_c)^2$ and 
$H_{c2}(T)\propto (1-T/T_c)^{1}$ in general agreement with 
experiments.
\end{abstract}


\vskip 0.5truecm 
\begin{center}
    To appear in {\it Physical Review Letters}
\end{center}

\narrowtext
\twocolumn

In the Ehrenfest classification of phase 
transitions\cite{pipp}, an $n^{th}$ order transition is described by 
continuous derivatives, with respect to temperature and a mechanical 
variable (for example a magnetic field or pressure) up to order 
$(n-1)$.  The $n^{th}$ order derivatives are discontinuous.  So far 
however, only first and second order transitions have been observed.  
There are no known examples of transitions higher in order than two.

We report below, what appears to be an example\cite{kum} of a $fourth$ 
order phase transition.  In the course of measuring\cite{hall} the 
magnetization of superconducting Ba$_{0.6}$K$_{0.4}$BiO$_3$ (BKBO) 
\cite{matt} , as a function of magnetic field (up to 27T) and 
temperature (1.3K to 350K) we were surprised to find no evidence of a 
discontinuity in the magnetic susceptibility.  While this was an 
anomalous property, it was congruent with the other mystery about 
BKBO, that there is no discontinuity in specific heat 
either\cite{hund} at $T_c$.  Since in a second order phase transition, 
the boundary between the normal and superconducting phases satisfies,

\beq
\left({dH_{c2}\over dT}\right)^2={\Delta C\over
T_c\Delta \chi},
\eeq  
with both $\Delta C$ and $\Delta \chi$ vanishing, a question arises 
concerning 
the order of this transition.

\begin{figure}[tbp]
	\centering
	\epsffile{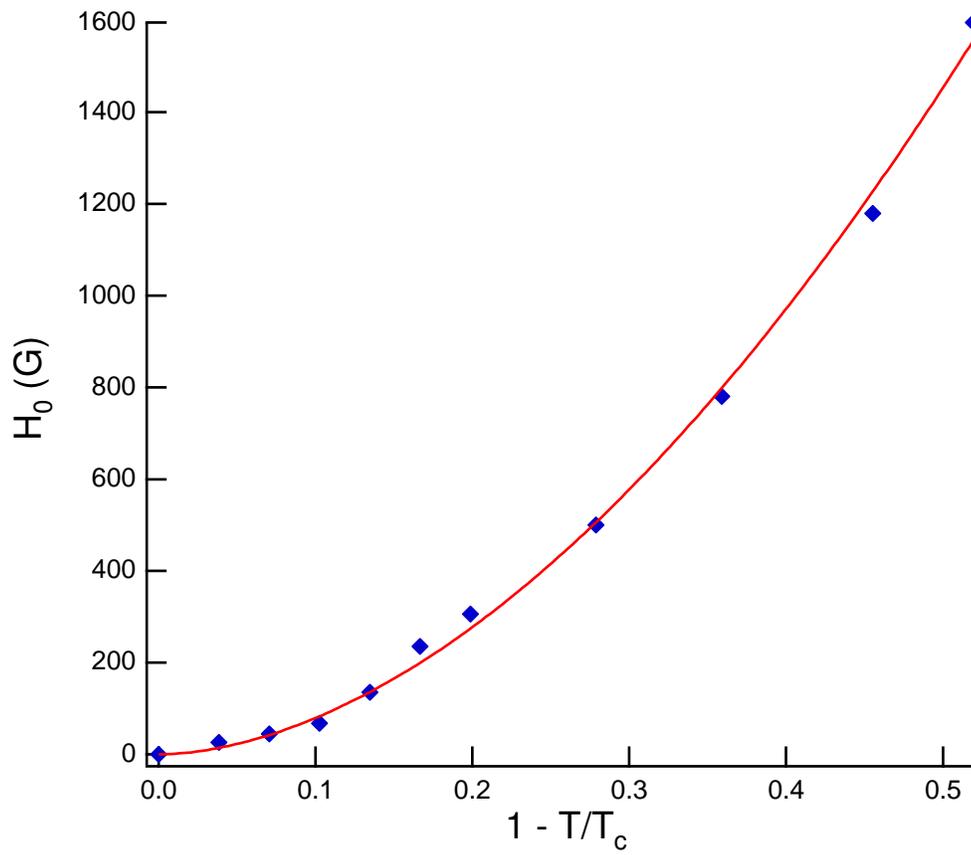}
	\caption{The values of the thermodynamic critical field $H_0$ are 
	plotted here as a function of $1-T/T_c$.   
	$H_{0}(T) = 0.509\cdot (1-T/T_c)^{1.807 \pm 0.052}$ tesla with $T_{c} = 32$ K.}
	\label{fig1}
\end{figure}

The answer is provided by the thermodynamic critical field $H_0(T)$.  
Since in the superconducting state\cite{tink}, the thermodynamic 
critical field, $H_{0}(T)$, is given by ($0<H<H_{c2}$), $\int 
{\bf M}\cdot d{\bf H}= - H_0^2/8\pi$.  \ This is the free energy of the 
superconducting state that is derived from the experimentally 
determined M(H,T); thus, in case of a second order phase transition, 
should have the temperature dependence $F(T) = - H_0^2/8\pi \propto - 
(1 - T/T_c)^2$.  \ That is, $H_0(T)$ would be linear in $(1-T/T_c)$, 
apart from critical fluctuation effects which lead to a divergent 
specific heat.  \ As shown in Fig. \ref{fig1}, with $T_{c} = 32$ K 
$H_0(T) \propto (1-T/T_c)^2$.\ Since 
for an $n^{th}$ order phase transition, the critical field has an 
exponent of $n/2$, the transition here must be of $fourth$ order in 
the sense of Ehrenfest.

\begin{figure}[tbp]
	\centering
	\epsffile{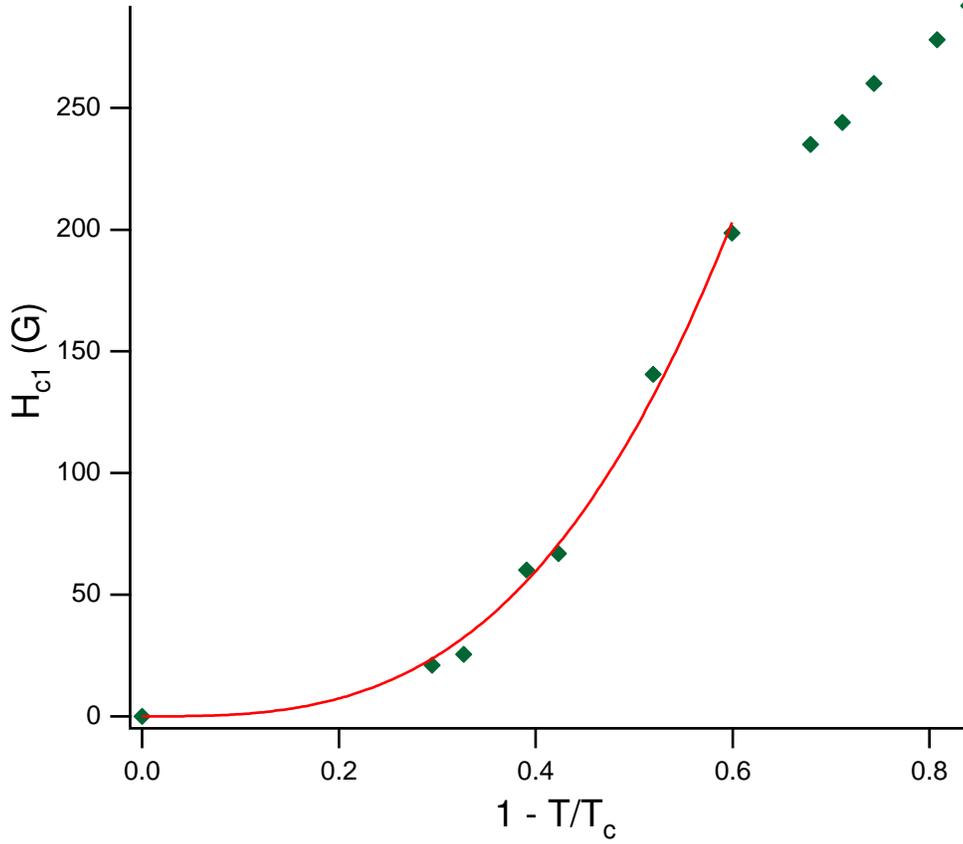} 
	\caption{The lower critical field 
	$H_{c1}$ is plotted as a function of $1-T/T_c$.  $H_{c1}(T) = 
	0.0955 \cdot (1-T/T_c)^{3.027 \pm 0.156}$ tesla with $T_{c} 
	= 32$ K. Below roughly $T = T_{c}/2$ the data is 
	approximately linear.}
	\label{fig2}
\end{figure}

\begin{figure}[tbp]
	\centering
	\epsffile{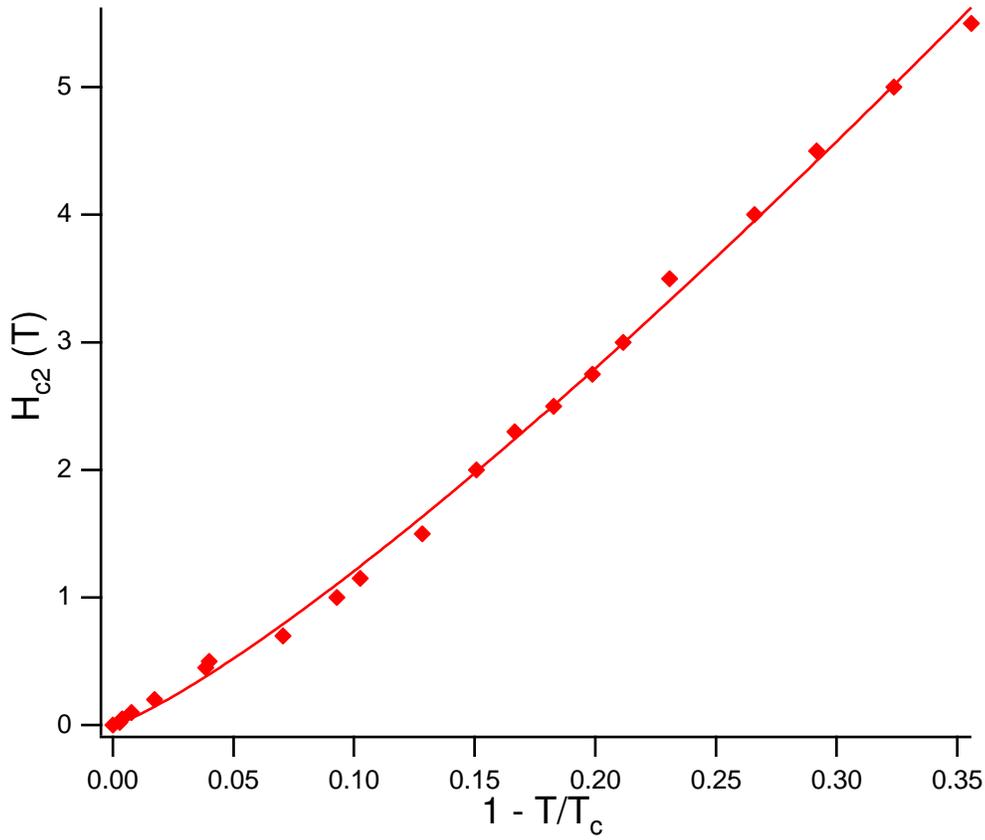}
	\caption{Shown here is the upper critical field $H_{c2}$ plotted as a function of $1-T/T_c$. 
	$H_{c2}(T) = 0.00197  \cdot (1-T/T_c)^{1.213 \pm 0.021}$ tesla with $T_{c} = 32$ K. }
	\label{fig3}
\end{figure}

Further support for this assertion comes from the temperature 
dependence of other critical fields.  In particular, we find experimentally
that the lower critical field, the field which separates the Meissner
state (no flux in the sample) from the Abrikosov state (partial flux penetration 
in the form of a vortex lattice), depends\cite{hall} on temperature as 
$H_{c1}(T)\propto (1-T/T_c)^3$ as shown in Fig. \ref{fig2}.  The upper 
critical field, which separates the Abrikosov state from the normal 
state is measured to be $H_{c2}(T) \propto (1-T/T_{c})^{1.2}$ as 
shown in Fig. \ref{fig3}.  This fact leads to an anomalous result, specific 
to this higher order phase transition.  For a BCS superconductor, the ratio 
${\kappa^2} = H_{c2}(T)/H_{c1}(T)$ is a constant.  Here
it diverges approximately as $(1-T/T_c)^{-2}$.  Both of the critical fields are 
inversely proportional to squares of the two length scales in the problem, 
the London penetration length $\lambda$, which controls the flux penetration 
and therefore the size of a vortex, and the superconducting coherence 
length $\xi$ which determines the stiffness of the local density of the 
superconducting electrons.  In a BCS superconductor, these length scales are 
identical in their temperature dependence.  To our knowledge there is no 
fundamental reason why $\kappa$ should be a constant. 

In the following we derive a free energy functional which describes the 
properties of a $fourth$ order phase transition.  Once we include the 
interaction of the superconductor with the magnetic field in the usual 
gauge invariant form, we also can  derive the temperature dependencies 
of the critical fields.  These results are in full accord with the 
experiments.

The free energy is derived following the requirement of a $fourth$ order
phase transition, viz. $F(T)= - f_o(1-T/T_c)^4$ as a function of 
temperature.  The free energy is in the spirit of a Ginzburg-Landau 
(GL)
functional which is minimized with respect to the complex order 
parameter $\psi=\Delta e^{i\phi}$.  The value at the minimum then 
is the thermodynamic free energy. The first two terms are self evident.  
Indeed it is important that the terms proportional to $|\psi|^2$ and
$|\psi|^4$ be not present.  The form of the spatial gradient
term is also determined by the same considerations.  The term below is 
the one with the lowest power of gradients.  Higher power of gradients 
such as $|\psi {\nabla}^2 \psi|^2$ and $|{\nabla}^3 \psi|^2$ are possible 
but they contribute higher order nonlinear contributions of the magnetic 
field and therefore are unnecessary for a stability analysis.  They can 
be included for effects nonlinear in the magnetic field.  The free 
energy functional appears as, 

\beq
F_{\rm IV} (\psi, T) = a |\psi|^6 + b|\psi|^8 + c|\psi^2
\nabla\psi|^2
\eeq
Here $a=a_o({T/T_c} - 1)$ and $b$ and $c$ are positive constants.

The minimum of this free energy corresponds to an order parameter amplitude 
$\Delta (T) \propto (1-T/T_c)^{1/2}$. The specific heat is expected to
be $C_{\rm IV}(T)\propto (1-T/T_c)^2$, and 
$\chi={\partial M\over \partial H}\propto (1-H/H_{c2})^2$. \ 
The thermodynamic discontinuities are in the fourth derivative of the free
energy (or in the second derivative of specific heat as a
function of temperature or the second derivative of the
magnetic susceptibility $\chi$ with respect to the magnetic
field).  We see that in the common thermodynamic observables,
there are no discontinuities, as seen in the experiments.  It is 
conceivable that broad transitions that have been observed in the past, instead of being recognized as 
candidates for a higher order phase transition were forcibly squeezed 
into a second order framework.  The Ehrenfest relation appropriate for 
a IV order phase transition is
\beq
\left( {dH\over dT}\right )^4 = {\Delta {\partial^2 c
\over \partial T^2}\over T_c \Delta {\partial^2 \chi
\over \partial H^2}}
\eeq

In the presence of a magnetic field the gradient term
transforms as $\nabla \rightarrow (\nabla + {2\pi i\over \phi_o} 
A)$, where A is the vector potential. \ Here $\phi_o$ is the flux quantum, 
$\phi_o =h/2e=2\times 10^{-15}T\cdot m^2$.
Thus Eq.~(2), as always, is the basis for a study of both spatial thermodynamic 
fluctuations as well as magnetic field effects.  We
note that the penetration depth for a magnetic field, 
the coefficient of the $A^2$ in the generalized Eq. (2), diverges as
\beq
\lambda^{-2}(T) = {4(2\pi)^3\over \phi_o^2} c\Delta^6
\propto (1-T/T_c)^3
\eeq
This too is in agreement with experiments, not as a direct
measurement but that of $H_{c1}(T) \propto \phi_o/\lambda^2$. 
This is shown in Fig. \ref{fig2} with $H_{c1}(T)$ plotted as a function of temperature.
Here the data on the lower critical field is limited by its size at 
temperatures close to $T_c$. At low temperatures (less than $T_c/2$), $H_{c1}(T)$
behaves linear in T and has the right intercept at $T_c$.
The spatial fluctuations of the order parameter are still
governed by $\xi^2=c/a \propto (1-T/T_c)^{-1}$. For $H_{c2}(T)$
we recognize that $\phi_o/\xi^2=H_{c2}(T)$.  Experimentally, 
as shown in Fig. \ref{fig3}, the exponent is nearly one.

The proposal here rests on several critical assumptions.  For example, 
Graebner et al\cite{grab} have reported a very small specific heat discontinuity.  
The reported discontinuity is in fact anomalously small and roughly of 
the size of their experimental uncertainly.  To estimate the expected\cite{kwok}
discontinuity, consider the specific heat results in ref\cite{hund}.  The high temperature limit 
of the specific heat can be described by ${C(T) = \gamma T + \beta T^2}$ 
with ${\gamma \simeq 150 mJ/mole K^2}$ as the electronic contribution 
to $C(T)$.  This large $\gamma$ puts BKBO in the category of heavy fermion 
compounds and the expected $\Delta C$ (of the order of $\gamma T_{c}$ should 
be nearly 5J/moleK, 
considerably more (by a factor of $10^5$) than the experimental uncertainty 
and the reported value in ref\cite{grab}.  Moreover, Hundley et al \cite{hund}
find that at low temperatures (${T < T_c/2}$), the linear term in $C(T)$ 
disappears. The specific heat then is given by $C= {{\beta}^{'}}T^3$, 
where $\beta^{`} > \beta$. But 
this larger ${\beta}^{'}$  may well be due to the presence of nodes in the 
putative energy gap at the Fermi surface. For example point nodes in 
the energy gap give rise to a $C \propto T^3$ augmenting the well known 
phonon contribution with the same power.

Another basis for the suggestion here is the temperature dependence of 
the lower critical field $H_{c1}(T)$.  The cubic temperature 
dependence here is in contrast to the results of Grader et 
al\cite{heb} where $H_{c1}(T)$ is linear, as expected for a second 
order BCS superconductor.  However, closer inspection reveals that the 
$H_{c1}(T)$ values of Hall et al\cite{hall} (the values used in this 
analysis) are at low temperatures ${T < T_c/2}$ in agreement with the 
results of Grader et al\cite{heb} who employed in their study high 
quality microcrystals to eliminate spurious effects associated with 
sample inhomogeneities.  The values given in ref \cite{heb} for ${T 
>T_c/2}$, while in general agreement with ref\cite{hall} can be seen 
to follow a straight line but extrapolate to a smaller $T_c \simeq 
27K$.  When the zero field $T_c \approx 32~K$ is included, it is 
impossible to avoid a curvature in the temperature dependence of 
$H_{c1}(T)$.

We note, in passing, that the relation ${H_o}^2 = H_{c1} H_{c2}$ is still valid.
The consequences, near $T_c$ of a divergent $\lambda$ are more curious.  For 
example the central result that the flux expulsion happens more slowly in 
the mixed state is clear.  That the vortex lattice appears more slowly and 
therefore the irreversibility field is smaller is less obvious. Other questions 
such as the symmetry of the vortex lattice are currently under study and will 
be reported later. Similarly, the surface energy of a 
normal-superconducting (N-S) domain wall,  is
negative and proportional to $\lambda$ and therefore larger than in a BCS case.
It may well engender a more inhomogeneous ground state at $H_{c2}$.  A 
numerical analysis of these questions is in progress and will be reported later. 

It is important to note that the thermodynamic behavior changes for 
$T\leq T_c/2$.  This is clearly seen in several independent 
measurements, for example specific heat and critical fields.  The 
discussion here is focussed on the order of the transition from the 
normal state and is therefore limited to the vicinity of $T_c$.  
However, a microscopic theory which might attempt to derive Eq.  (2) 
will also have to include an explanation of this crossover behavior 
and possible existence of point nodes in the energy gap.

It might also contain an explanation of why the free energy does not
contain terms such as $\Psi^2$ and $\Psi^4$.  At present we can only
speculate about a microscopic theory.  In a sense, this question is
equivalent to the seemingly deeper question: Why is the transition of
order IV? In this paper, we have focussed on the properties of a IV order
phase transition, but let's speculate:  for instance, the BCS/GL theory
contains an overall factor of density of states at the Fermi surface.
Suppose, as discussed in ref. \cite{kum}, the density of states $N(0)=0$ for
${T\geq T_c}$ and $N(0) \propto {\Psi^{2p}}$ for $T\leq T_c$.  This would be a
transition from an insulator to a superconductor, the free energy for
$p=1$ would not have a $\Psi^2$ term and the order of the transition 
would be III.  For p=2, the transition would be of order IV.

Now, addressing the relationship between
fluctuations as developed for a II order phase transition and the
framework: in a second order phase transition, including the
critical effects, one might view the free energy as depending on
temperature as $F_o(T)=-f_o(1-T/T_c)^{(2-\alpha)}$  Thus the small quantity
$\alpha$ is calculated by pseudo-perturbative schemes (such as Gaussian
approximation or some version of renormalization group).  It is clear,
however, that a value of $\alpha=-1\: or\: -2$ is essentially beyond the realm
of a perturbative approach.  If the free energy exponent is
significantly different from 2, then the unperturbed ground state could be
a transition of order corresponding to the nearest integer, about which 
a calculation of fluctuations could be done in the future.

The conclusions presented here are the first part of a work in 
progress.  We are currently working on determining (1) the magnitude 
of the fluctuations, and (2) whether there is an upper critical 
dimension and, if so, what it is.  These, and other points of 
interest, will be presented in forthcoming publications.

In summary then, we have analyzed the thermodynamic properties of the 
superconducting phase transition in Ba$_{0.6}$K$_{0.4}$BiO$_3$ (BKBO).  
The absence of a discontinuity in specific heat and magnetic 
susceptibility, on transforming from the normal to superconducting 
state, shows that the phase transition cannot be of second order.  The 
temperature dependence of the thermodynamic critical field shows that 
the transition is $fourth$ order.  The conclusions about other 
critical fields, derived from a free energy developed for a fourth 
order phase transition, i.e. $H_{c1}(T) \propto (1-T/T_c)^3$ and 
$H_{c2}(T) \propto (1-T/T_c)$ are in accord with the experiments.

The work at LSU was supported by NSF grant No. DMR-9501419.  We also acknowledge 
the support (in many ways) of the National High Magnetic Field Laboratory which is 
supported by the NSF Cooperative Agreement No. DMR 9016241. Significant part of the 
writing of this paper was done during a stay at the Theory Division at the 
Los Alamos National Laboratory.  PK is grateful to Alan Bishop for his hospitality
and the members of the T-11 group for discussions.  We thank M. Graf, A. Hebard, Daryl 
Hess, J. R. Schrieffer, J. D. Thompson and J. W. Wilkins for their insights 
and discussion.

\newpage

\onecolumn
\widetext

\newpage

\newpage

\end{document}